\documentclass{revtex4}
\usepackage{graphicx}
\usepackage{fancyhdr}
\usepackage{amsmath}
\usepackage{color}

\newcommand{\GeV}{\mbox{GeV}}

\newcommand{\bequ}{\begin{equation}}
\newcommand{\eequ}{\end{equation}}
\newcommand{\beqn}{\begin{eqnarray}}
\newcommand{\eeqn}{\end{eqnarray}}
\newcommand{\bctr}{\begin{center}}
\newcommand{\ectr}{\end{center}}
\newcommand{\bit}{\begin{itemize}}
\newcommand{\eit}{\end{itemize}}

\def\lromn#1{\uppercase\expandafter{\romannumeral#1}}

\def\GeV{\rm{GeV}}

\begin{document}

\title{Higgs Phenomenology as a Probe of Supersymmetric Grand
  Unification with the Hosotani Mechanism\footnote{Talk presented at
    the International Workshop on Future Linear Colliders (LCWS13),
    Tokyo, Japan, 11-15 November 2013.  This talk is based on the work
    in collaboration with Shinya Kanemura, Hiroyuki Taniguchi and
    Toshifumi Yamashita \cite{SGGHU}.}}
  \author{Mitsuru Kakizaki}
  \affiliation{
  Department of Physics,
  University of Toyama, Toyama 930-8555, Japan
  }
  \begin{abstract}
    In the supersymmetric $SU(5)$ grand unified theory whose gauge
    symmetry is broken by virtue of the Hosotani mechanism, the huge
    mass splitting between the colored Higgs triplet and the
    electroweak Higgs doublet superfields is naturally realized.  As a
    byproduct, the existence of adjoint chiral superfields with masses
    of the order of the supersymmetry breaking scale is predicted, leading
    to the Higgs sector that
    contains an $SU(2)_L$ triplet chiral multiplet with hypercharge
    zero and a neutral singlet one in addition to the two $SU(2)_L$
    doublets of the minimal supersymmetric standard model.  We focus
    on the Higgs sector and investigate to what extent the couplings
    of the standard model-like Higgs boson and the masses of the
    additional Higgs bosons deviate from those in the Standard Model
    and other models due to the new triplet and singlet chiral
    multiplets.  We show that this model can be distinguished using
    precision measurements of couplings and masses of the Higgs sector
    particles and serves as a good example of grand unification
    testable at colliders such as the luminosity up-graded Large
    Hadron Collider and future electron-positron colliders.
  \end{abstract}
\maketitle

\thispagestyle{fancy}

\section{Introduction}

In 2012, the ATLAS and CMS collaborations at the CERN Large Hadron
Collider (LHC) discovered a new particle whose mass is approximately
125 GeV \cite{LHC}.  Its spin and CP properties as well as its
couplings with other particles have been analyzed.  So far, no clear
evidence that contradicts the properties of the Standard Model (SM)
Higgs boson has been found.  The SM is now confirmed as a low energy
effective theory that successfully explains phenomena below the TeV
scale.  

The SM, however, bears problems and puzzles that should be solved in a
more fundamental theory.  These include the hierarchy problem and the
charge quantization problem.  To keep the mass of the Higgs boson to
the electroweak scale, an unnatural huge cancellation between its bare
mas squared and quadratically divergent contributions from radiative
corrections is required.  Although the electric charges of the SM
particles can be theoretically arbitrary, they must be fractionally
quantized to account for the neutrality of atoms.

It is a natural idea to employ larger symmetries to tackle such
problems.  Indeed, the above-mentioned problems can be resolved by
supersymmetry (SUSY) and grand unification \cite{GUT,SUSY-GUT}.  In
supersymmetric extensions of the SM, the quadratically divergent
contributions from SM particles to the Higgs boson mass squared are
canceled with those from their partner particles, and therefore the
hierarchy problem is avoided.  Grand Unified Theories (GUTs) give a
unified description of the three SM gauge groups and SM fermions.  If
the gauge group of a GUT is (semi-)simple, the electric charges of the
SM particles are automatically quantized.  Therefore, models that have
both SUSY and a grand unified symmetry are well-motivated candidates
for physics beyond the SM.

Although SUSY GUT models contain such appealing features, there are
several unattractive points.  The typical GUT scale where the three
gauge couplings are unified is around $10^{16}$ GeV in usual SUSY
GUTs.  According to the decoupling theorem, the effects of the heavy
GUT particles are negligible at the TeV scale \cite{Appelquist:1974tg}.
One can obtain information on physics realized at the GUT scale only
through the relations among the mass and coupling parameters measured
at the TeV scale.  In addition, an unnaturally huge mass splitting
between the colored Higgs triplets and the electroweak Higgs doublets
are assumed to suppress the proton decay rate adequately.  Many
mechanisms have been proposed to solve this doublet-triplet splitting
problem
\cite{DW,SlidingSinglet,missingPARTN,pNG,orbifoldGUTs,Kakizaki:2001en}.

Recently, a SUSY GUT model that predicts the existence of new
particles accessible at collider experiments is proposed
\cite{gGHU-DTS}.  In this model, the doublet-triplet mass splitting is
naturally realized by supersymmetrizing the Grand Gauge-Higgs
Unification (GHU) model \cite{gGHU}.  The Supersymmetric Grand
Gauge-Higgs Unification (SGGHU) is constructed on an extra dimension
whose compactification scale is around $10^{16}$ GeV, where the GUT
gauge symmetry is broken by the Hosotani mechanism \cite{hosotani}:
The non-trivial vacuum expectation value (VEV) of the fifth component
of one of the gauge fields is responsible for the symmetry breaking.
In the SGGHU model, a color octet superfield, an $SU(2)_L$ triplet
superfield with hypercharge zero and a neutral singlet superfield
appear at the TeV scale as a by-product.  As compared to the minimal
supersymmetric standard model (MSSM), the Higgs sector is extended by
the triplet and singlet superfields.

In this talk, we discuss the properties of the SGGHU Higgs sector, and
investigate its phenomenological signatures expected at collider
experiments.  We evaluate the masses and couplings of the Higgs sector
particles solving renormalization group equation (RGE) from the GUT
scale to the electroweak scale.  Our emphasis is that precision
measurements of the masses and couplings of the Higgs bosons at the
LHC and future electron-positron colliders such as the International
Linear Collider (ILC) \cite{ILC} and the CLIC \cite{CLIC} play an
important role in distinguishing particle physics models.  The SGGHU
model is a good example to show that low-energy collider experiments
are capable of testing GUT scale physics.

\section{Model of Supersymmetric Grand Gauge-Higgs Unification}
\label{Sec:Model}

Here, we briefly discuss the structure of the Higgs sector of the low
energy effective theory of the SGGHU model, which contains an $SU(2)_L$
triplet chiral superfield $\Delta$ and an neutral singlet chiral
superfield $S$ as well as the two MSSM Higgs doublets $H_u$ and $H_d$.
The superpotential of the SGGHU Higgs sector is
given by
\begin{eqnarray} \label{eq:WHiggs}
  W=\mu H_u \cdot H_d+\mu_\Delta^{}{\rm tr}(\Delta^2)+\frac{\mu_S^{}}2S^2
  +\lambda_\Delta^{} H_u \cdot \Delta H_d + \lambda_S^{}SH_u \cdot H_d \, ,
\end{eqnarray}
where $\Delta = \Delta^a \sigma^a/2$ with $\sigma^a (a=1,2,3)$ being the
Pauli matrices.  The fact that $S$ and $\Delta$ stem from the gauge
supermultiplet leads to the following remarkable features.  Although
trilinear self-couplings among $S$ and $\Delta$ are not forbidden in
the general Higgs superpotential that contains the triplet and singlet
superfields, such terms are absent in our model.  The newly introduced
Higgs couplings $\lambda_\Delta^{}$ and $\lambda_S^{}$ are unified
with the SM gauge coupling constants at the GUT scale.  Therefore,
this model can predict the properties of the Higgs bosons with less
ambiguity.  The soft SUSY breaking terms in the Higgs potential read
\begin{eqnarray} \label{eq:VsoftHiggs}
  V_{\rm soft} 
  &=& \widetilde{m}_{H_d}^2 |H_d|^2 +\widetilde{m}_{H_u}^2|H_u|^2
  + 2 \widetilde{m}_\Delta^2 {\rm tr} (\Delta^\dag \Delta)
  +\widetilde{m}_S^2|S|^2 \nonumber \\
  &&+\left[B\mu H_u\cdot H_d +\xi S 
    + B_\Delta^{} \mu_\Delta^{}{\rm tr}(\Delta^2) +\frac{1}{2}B_S^{} \mu_S^{} S^2 
  +\lambda_\Delta^{}A_\Delta^{} H_u\cdot \Delta H_d +\lambda_S^{} A_S^{} S H_u\cdot H_d +{\rm h.c.} \phantom{\frac{1}{2}}\right].
\end{eqnarray}
The values of the soft parameters at the TeV scale are also determined
by solving the RGEs.  Due to the top Yukawa contributions to the RGEs,
radiative electroweak symmetry breaking occurs.  Then, in the Higgs
sector, four CP-even, three CP-odd and three charged Higgs bosons
appear as physical particles.  The VEV of the neutral component of the
triplet Higgs boson $v_\Delta^{}$ is obtained from the minimization
conditions of the Higgs potential, and must be less than $\simeq
10~{\rm GeV}$ to satisfy the rho parameter constraint.  Since this
value is sufficiently small compared to $v=246~{\rm GeV}$, $v_\Delta$
is neglected in our computation of the Higgs boson masses and
couplings.

\section{Analysis of  the Renormalization Group Equations}
\label{Sec:RGEAnalysis}

\begin{figure}[t]
\centering
\includegraphics[width=90mm]{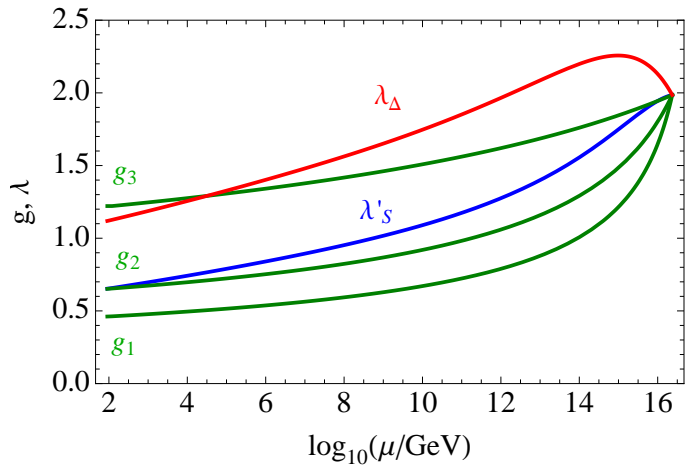}
\caption{\footnotesize Evolution of the Higgs triplet and
  singlet coupling constants $\lambda_\Delta^{}$ (red line) and
  $\lambda'_S$ (blue) as well as the gauge coupling constants $g_3$,
  $g_2$ and $g_1$ (green) at the one loop level 
  as a function of the energy scale.}
\label{fig:couplings}
\end{figure}

Let us turn to discussions about RG evolution of the coupling
constants and mass parameters in the SGGHU model.  As a consequence of
introducing the light adjoint multiplets, the successful gauge
coupling unification is disturbed.  In our model, by adding extra
incomplete $SU(5)$ matter multiplets, the gauge coupling unification
can be easily recovered.  An successful example for the matter
multiplets is a set of two vectorlike pairs of $(\bar{L},L)$
$(({\bf1},{\bf2})_{-1/2})$, one of $(\bar{U},U)$ $((\bar
{\bf3},{\bf1})_{-2/3})$ and one of $(\bar{E},E)$
$(({\bf1},{\bf1})_{1})$, where the numbers in the parentheses are
$SU(3)_C$, $SU(2)_L$ and $U(1)_Y$ quantum numbers, respectively
\cite{gGHU-DTS}.  Fig.~\ref{fig:couplings} shows the evolution of the
Higgs triplet and singlet coupling constants $\lambda_\Delta^{}$ (red
line) and $\lambda_S^{}$ (blue), as well as the gauge coupling
constants $g_3^{}$ , $g_2^{}$ and $g_1^{}$ (green) at the one loop
level as a function of the energy scale.  The coupling constants are
normalized such that $\lambda'_S=(2\sqrt{5/3}) \lambda_S^{}$ for the
singlet Higgs coupling and $g_1=(\sqrt{5/3})g_Y$ for the $U(1)_Y$
gauge coupling.  The Higgs trilinear coupling constants at the
TeV-scale are
\begin{equation} \label{eq:couplings}
  \lambda_\Delta^{}=1.1\, , \qquad \lambda_S^{}=0.25\, .
\end{equation}
The soft breaking parameters at the SUSY breaking scale are evaluated
by solving the RGEs.  Fig.~\ref{fig:couplings} shows that we have
strong gauge couplings at the GUT scale.  To satisfy the gluino mass
limit \cite{LHCgluino}, the unified gaugino mass at the GUT scale
needs to be large.  Consequently, soft sfermion and Higgs masses at
the SUSY breaking scale are typically multi-TeV.  This means that one
needs some tuning to have successful radiative electroweak symmetry
breaking.  In spite of the difficulties, we will show that one can also
obtain soft Higgs masses of the order of ${\cal O}(100)$ GeV by tuning
among the input parameters at the GUT scale.

\section{Impact on Higgs properties}

First, let us discuss the prediction about the SM-like Higgs boson mass, 
which is affected by the $SU(2)_L$ triplet and singlet Higgs multiplets.
When the soft scalar masses of the triplet and singlet are relatively large,
the mass of the SM-like Higgs boson is approximately given by
\cite{MSSMHiggsMass,Espinosa}
\begin{align} \label{eq:higgs_mass}
m_h^2 \simeq
m_Z^2 \cos^2 \beta
+\frac{3 m_t^4}{2\pi^2 v^2}\left( \ln{\frac{m_{\tilde{t}}^2}{m_t^2}} +
  \frac{X_t^2}{m_{\tilde{t}}^2} \left(1 -\frac{X_t^2}{12m_{\tilde{t}}^2} 
    \right) \right)
+\frac{1}{8}\lambda_\Delta^2 v^2 \sin^2 {2\beta}
+\frac{1}{2}\lambda_S^2 v^2 \sin^2 {2\beta}
\, ,
\end{align}
where $m_Z$ is the $Z$-boson mass, $m_t$ is the top quark mass,
$m_{\tilde{t}}$ is the averaged stop mass, and $X_t = A_t - \mu \cot
\beta$.  In the MSSM, for the SM-like Higgs boson mass to reach
$125~\GeV$ using the large stop loop contribution, one needs very
large stop masses even in the maximal stop mixing scenario
\cite{MSSMHiggsMass2}.  As in the next-to-MSSM (NMSSM) \cite{NMSSM},
in our model the SM-like Higgs boson mass is lifted up by the Higgs
trilinear couplings with the triplet and singlet superfield, in
particular, for small $\tan \beta$ region.  In our numerical
computation of the masses of the Higgs scalars and superparticle, we
have used the public numerical code \texttt{SuSpect} \cite{suspect}
after including the contributions from the triplet and singlet Higgs
superfields, instead of the approximate formula
Eq.(\ref{eq:higgs_mass}).
Since fine tuning for the GUT-scale input parameters
is required, we show results based on some benchmark points
that reproduce the correct SM-like Higgs boson mass.
Taking theoretical uncertainties into account,
we allow the mass range $122~{\rm GeV} < m_h < 129~{\rm GeV}$.
We consider the following three typical scenarios:
\begin{itemize}
\item[(A)] Mixings between the SM-like Higgs boson and the other Higgs bosons are small.
\item[(B)] Mixings between the SM-like Higgs boson and the triplet and singlet Higgs bosons are small.
\item[(C)] The triplet and singlet Higgs bosons affect the SM-like Higgs boson couplings.
\end{itemize}
Three successful benchmark points for the GUT-scale input parameters and the
resulting TeV-scale parameters obtained after RG evolution are shown in
Tab. \ref{tab:benchmark-GUT} and \ref{tab:benchmark-TeV}, respectively.

\begin{table}[t]
  \begin{flushleft}
{\footnotesize
  \begin{tabular}{|c||c|c|c|c|c|c|c|}
    \hline
    Case &
    $\tan \beta$ &
    $M_{1/2}$ &
    $\mu_{\Sigma}$
    \\ \hline  \hline
    (A)(B)(C) &
    $3$ &
    $3600~{\rm GeV}$ &
    $-300~{\rm GeV}$
    \\ \hline 
  \end{tabular}
  \begin{tabular}{|c||c|c|c|c|c|c|c|}
    \hline
    Case &
    $A_0$ & 
    $\widetilde{m}_0^2$ & 
    $\widetilde{m}_{H_u}^2$ &
    $\widetilde{m}_{H_d}^2$ &
    $\widetilde{m}_{5}^2$ &
    $\widetilde{m}_{10}^2$ &
    $\widetilde{m}_\Sigma^2$
    \\ \hline  \hline
    (A) &
    $5500~{\rm GeV}$ & 
    $(1000~{\rm GeV})^2$ &
    $(10375~{\rm GeV})^2$ &
    $(8570~{\rm GeV})^2$ & 
    $- (6300~{\rm GeV})^2$ &
    $- (2000~{\rm GeV})^2$ &
    $- (570~{\rm GeV})^2$
    \\ \hline 
    (B) &
    $1000~{\rm GeV}$ & 
    $(1800~{\rm GeV})^2$ &
    $(12604~{\rm GeV})^2$ &
    $(10381.5~{\rm GeV})^2$ & 
    $- (7700~{\rm GeV})^2$ &
    $- (1960~{\rm GeV})^2$ &
    $- (670~{\rm GeV})^2$
    \\ \hline 
    (C) &
    $8000~{\rm GeV}$ & 
    $(3000~{\rm GeV})^2$ &
    $(10605.1~{\rm GeV})^2$ &
    $(8751.4~{\rm GeV})^2$ & 
    $- (6418~{\rm GeV})^2$ &
    $- (1638.5~{\rm GeV})^2$ &
    $- (400~{\rm GeV})^2$
        \\ \hline 
  \end{tabular}
}
\end{flushleft}
  \caption{\footnotesize Benchmark points for the GUT-scale input parameters.}
  \label{tab:benchmark-GUT}
\end{table}

\begin{table}[t]
  \begin{flushleft}
    {\footnotesize
  \begin{tabular}{|c||c|c|c|c|c|c|c|c|}
    \hline
    Case &
    $M_1$ &
    $M_2$ &
    $M_3$ &
    $\mu_{\Delta}$ &
    $\mu_S^{}$
    \\ \hline  \hline
    (A)(B)(C) &
    $194~{\rm GeV}$ & 
    $388~{\rm GeV}$ &
    $1360~{\rm GeV}$ &
    $-252~{\rm GeV}$ &
    $-85.8~{\rm GeV}$
    \\ \hline 
\end{tabular}
  \begin{tabular}{|c||c|c|c|c|c|c|c|c|}
    \hline
    Case &
    $\mu$ &
    $B \mu$ & 
    $\widetilde{m}_{u_3}$ & 
    $\widetilde{m}_{q_3}$ &
    $y_t A_t$
    \\ \hline  \hline
    (A) &
    $205~{\rm GeV}$ &
    $41400~{\rm GeV}^2$ &
    $3290~{\rm GeV}$ & 
    $4830~{\rm GeV}$ &
    $4030~{\rm GeV}$
    \\ \hline 
    (B) &
    $177~{\rm GeV}$ &
    $40800~{\rm GeV}^2$ &
    $1730~{\rm GeV}$ & 
    $4480~{\rm GeV}$ &
    $6050~{\rm GeV}$
    \\ \hline 
    (C) &
    $174~{\rm GeV}$ &
    $42000~{\rm GeV}^2$ &
    $4220~{\rm GeV}$ & 
    $5550~{\rm GeV}$ &
    $2910~{\rm GeV}$
    \\ \hline 
  \end{tabular}
  \begin{tabular}{|c||c|c|c|c|c|c||c|}
    \hline
    Case &
    $\widetilde{m}_{\Delta}$ & 
    $\widetilde{m}_{S}$ &
    $\lambda_\Delta^{} A_\Delta^{}$ &
    $\lambda'_S A_S^{}$ &
    $B_\Delta^{} \mu_\Delta^{}$ &
    $B_S^{} \mu_S^{}$ &
    $m_h$
    \\ \hline  \hline
    (A) &
    $607~{\rm GeV}$ & 
    $805~{\rm GeV}$ &
    $662~{\rm GeV}$ &
    $683~{\rm GeV}$ &
    $92000~{\rm GeV}^2$ &
    $-78700~{\rm GeV}^2$ &
    $123~{\rm GeV}$
    \\ \hline 
    (B) &
    $784~{\rm GeV}$ & 
    $612~{\rm GeV}$ &
    $1340~{\rm GeV}$ &
    $1110~{\rm GeV}$ &
    $30700~{\rm GeV}^2$ &
    $-110000~{\rm GeV}^2$ &
    $123~{\rm GeV}$
    \\ \hline 
    (C) &
    $521~{\rm GeV}$ & 
    $216~{\rm GeV}$ &
    $284~{\rm GeV}$ &
    $446~{\rm GeV}$ &
    $207000~{\rm GeV}^2$ &
    $-33600~{\rm GeV}^2$ &
    $122~{\rm GeV}$
    \\ \hline 
  \end{tabular}
}
  \end{flushleft}
  \caption{\footnotesize TeV-scale parameters obtained after RG running.}
  \label{tab:benchmark-TeV}
\end{table}

\begin{figure}[t]
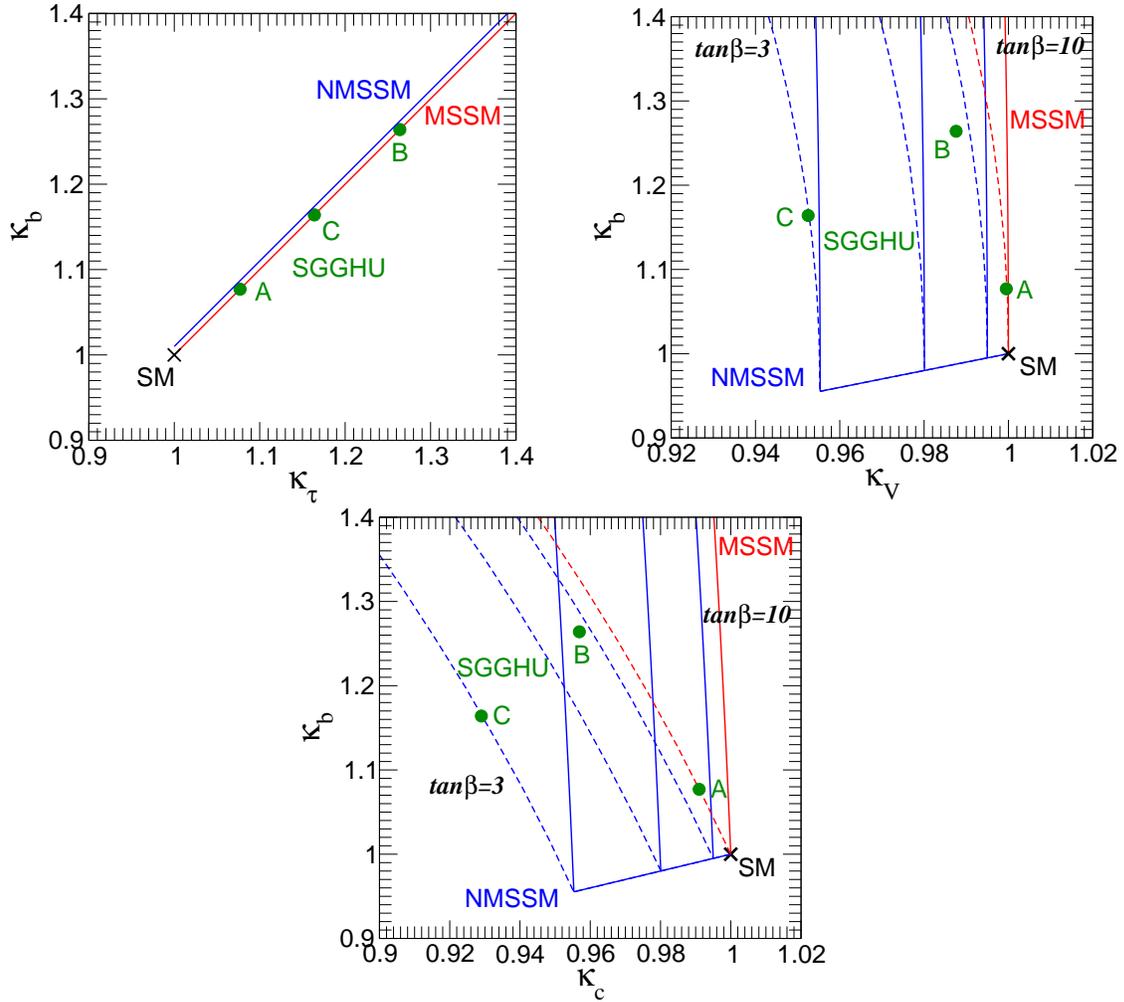

  \begin{center}
      \begin{minipage}{0.45\hsize}
        \begin{center}
          \includegraphics[clip,width=70mm]{taub.eps}    
        \end{center}
      \end{minipage}
      \begin{minipage}{0.45\hsize}
        \begin{center}
          \includegraphics[clip,width=70mm]{vb.eps}    
        \end{center}
      \end{minipage}
      \begin{minipage}{0.45\hsize}
        \begin{center}
          \includegraphics[clip,width=70mm]{cb.eps}    
        \end{center}
      \end{minipage}

      \caption{\footnotesize The deviations in the scaling factors
        $\kappa_X^{}$ are plotted on the
        $\kappa_\tau^{}$-$\kappa_b^{}$ plane, the
        $\kappa_V^{}$-$\kappa_b^{}$ plane $(V=Z,W)$, and the
        $\kappa_c^{}$-$\kappa_b^{}$ plane.  The deviations in the
        three benchmark scenarios (A), (B) and (C) in the SGGHU are
        shown with green blobs.  The MSSM predictions are shown with
        red lines for $\tan \beta=10$ (thick line) and $\tan \beta=3$
        (dashed).  The NMSSM predictions are shown with blue grid
        lines for $\tan \beta=10$ (thick) and $\tan \beta=3$ (dashed),
        which indicate mixings between the SM-like and singlet-like
        Higgs bosons of 10\%, 20\% and 30\% from the right to the
        left. }
    \label{fig:fingerprint}
  \end{center}
\end{figure}

The couplings between the SM-like Higgs boson and SM particles can be
significantly affected by the existence of the new Higgs bosons.  In
fingerprinting the SM-like Higgs boson couplings, it is useful to
define the following scaling factors
\begin{eqnarray}
  \kappa_X^{} = \frac{g_{hXX}^{}}{g_{hXX}^{}|_{\rm SM}^{}} \, ,
\end{eqnarray}
where $g_{hXX}^{}$ denotes the coupling with the SM particle $X$.  In
Fig.~\ref{fig:fingerprint}, the deviations in the scaling factors
$\kappa_X^{}$ are plotted on the $\kappa_\tau^{}$-$\kappa_b^{}$ plane,
the $\kappa_V^{}$-$\kappa_b^{}$ plane $(V=Z,W)$ and the
$\kappa_c^{}$-$\kappa_b^{}$ plane \footnote{The plots shown in
  Fig.~\ref{fig:fingerprint} are updated from the ones presented at
  LCWS13.}.  The deviations in the three benchmark scenarios (A), (B)
and (C) in the SGGHU are shown with green blobs.  The MSSM predictions
are shown with red lines for $\tan \beta=10$ (thick line) and $\tan
\beta=3$ (dashed).  The NMSSM predictions are shown with blue grid
lines for $\tan \beta=10$ (thick) and $\tan \beta=3$ (dashed), which
indicate mixings between the SM-like and singlet-like Higgs bosons of
10\%, 20\% and 30\% from the right to the left.  Since the Higgs boson
couplings to the down-type quarks and charged leptons are common in
our model as in the Type-II two Higgs doublet model, the predicted
SGGHU deviations lie on the MSSM and NMSSM lines on the
$\kappa_\tau^{}$-$\kappa_b^{}$ plane.  At the ILC with
$\sqrt{s}=500~{\rm GeV}$, expected accuracies for the scaling factors
$\kappa_Z$, $\kappa_W$, $\kappa_b^{}$, $\kappa_\tau^{}$ and
$\kappa_c^{}$ are 1.0\% 1.1\% 1.6\%, 2.3\% and 2.8\%, respectively
\cite{ILCHiggs}.  These plots show that characteristic SGGHU
predictions about the Higgs couplings are distinguishable from those
of the SM and MSSM by measuring the Higgs boson couplings accurately
at the future ILC while it may be difficult to completely distinguish
our model from the NMSSM only through the precision measurements.
Nevertheless, if the pattern of the Higgs coupling deviations turns
out to be close to one of our benchmark points, the possibility of the
SGGHU is increased.  Independent measurement of $\tan \beta$ utilizing
Higgs boson decay at the ILC \cite{Gunion:2002ip,Kanemura:2013eja}
will be also useful in distinguishing new physics models.  As for the
Higgs coupling with the photon and Higgs self-coupling, we obtained
$0.94 < \kappa_\gamma^{} < 1.0$ and $0.82 < \kappa_h^{} < 0.93$,
respectively, for the above benchmark points.  To observe such
deviations one needs more precise measurements at the ILC with
$\sqrt{s}=1~{\rm TeV}$ \cite{ILCHiggs}.

\begin{figure}[t]
\includegraphics[width=90mm]{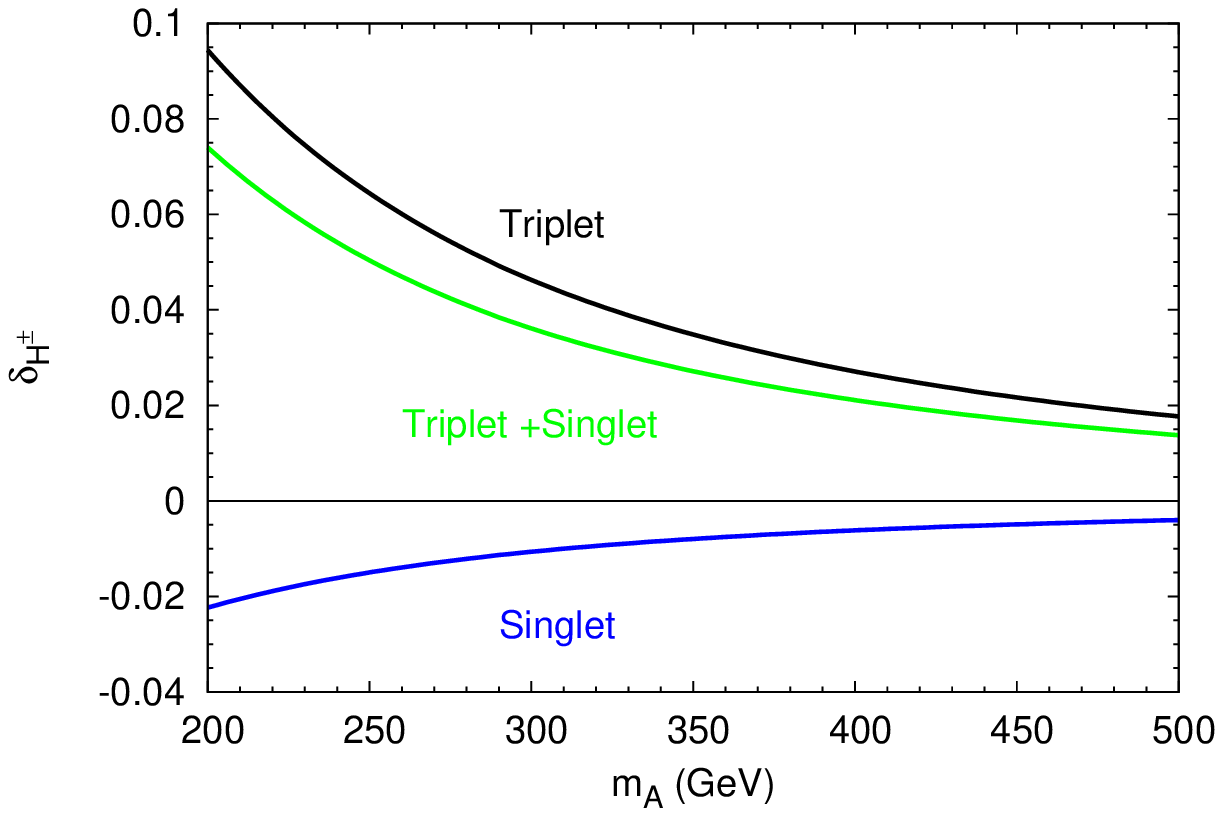}
\caption{\footnotesize The deviation parameter $\delta_{H^\pm}^{}$ of the
  MSSM-like charged Higgs boson mass as a function of the
  MSSM-like CP-odd Higgs boson mass $m_A^{}$ for relatively large
  soft Higgs masses.  The black, blue and green lines correspond to triplet
  contribution, singlet contribution and their sum, respectively.}
\label{fig:dhpm_ma}
\end{figure}

Let us mention the masses of the additional MSSM-like Higgs
bosons.  For relatively large soft scalar masses of the triplet
and singlet, the MSSM-like charged Higgs boson mass $m_{H^\pm}^{}$ is
approximately written as
\begin{align} 
  m_{H^\pm}^2
  = m_{H^\pm}^2|_{\rm MSSM}^{} (1 +\delta_{H^\pm}^{})^2 
  \simeq m_A^2 +m_W^2 +\frac{1}{8}\lambda_\Delta^2 v^2
  -\frac{1}{2}\lambda_S^2v^2\, ,
\end{align}
where $m_A$ is the MSSM-like CP-odd Higgs boson mass, and
$\delta_{H^\pm}^{}$ parametrizes the deviation of $m_{H^\pm}^{}$ from
the MSSM prediction.  The sign difference between the triplet and
singlet contributions comes from group theoretical factors.  Since
$\lambda_\Delta$ is significantly larger than $\lambda_S$ due to
radiative corrections, the MSSM-like charged Higgs boson mass in our
model is larger than the MSSM prediction.  Fig.~\ref{fig:dhpm_ma}
shows the deviation parameter $\delta_{H^\pm}^{}$ of the MSSM-like
charged Higgs boson mass as a function of $m_A^{}$ for relatively large soft
Higgs masses.  The black, blue and green lines correspond to triplet
contribution, singlet contribution and their sum, respectively.
When the MSSM-like Higgs bosons are lighter than $500~{\rm GeV}$,  
the mass deviation is found to be $\delta_{H^\pm}^{} \sim O(1)~\%$ - $O(10)~\%$ 
and detectable at the LHC \cite{HiggsWG}.

If the triplet-like and singlet-like Higgs bosons are lighter than
$500~{\rm GeV}$, the new Higgs bosons can be directly produced at the
ILC and CLIC.  As shown in Tab.~\ref{tb:spectrum}, the benchmark point
(C) contains such light Higgs bosons.  For example, $\Delta^{\pm}$ can
be probed through the channel $e^+ e^- \to \Delta^+ \Delta^- \to
t\bar{b} \bar{t} b$, which is induced by the mixing between the
MSSM-like and triplet-like charged Higgs bosons.

\begin{table}[t]
  \begin{tabular}{|c|c|c|}
    \hline
    CP-even & CP-odd & Charged \\ \hline \hline
    $122~{\rm GeV}$     & $-$                & $-$         \\ \hline
    $139~{\rm GeV}$     & $171~{\rm GeV}$    & $204~{\rm GeV}$     \\ \hline
    $370~{\rm GeV}$     & $304~{\rm GeV}$    & $496~{\rm GeV}$     \\ \hline
    $745~{\rm GeV}$     & $497~{\rm GeV}$    & $745~{\rm GeV}$     \\ 
    \hline
  \end{tabular}
  \caption{Higgs boson masses for the benchmark point (C).} 
  \label{tb:spectrum}
\end{table}

Above discussions show that new physics models can be distinguished
through exhaustive analysis of the masses and couplings of the Higgs
sector particles at the LHC and future electron-positron colliders.
Even when the additional Higgs bosons are beyond the reach of direct
discovery, their effects can be indirectly probed by precise
measurements of the SM-like Higgs boson couplings and the MSSM Higgs
boson masses.  A new electron-positron collider is mandatory
for exploring the Higgs properties and the underlying theory.

\section{Summary}

In the SUSY $SU(5)$ GUT model where the Hosotani mechanism is
responsible for the $SU(5)$ gauge symmetry breaking, the Higgs sector
at the TeV scale contains a Higgs triplet and singlet chiral
superfields as well as the two MSSM Higgs doublets.  We have computed
the couplings between the SM-like Higgs boson and SM particles.  The
deviations of these coupling constants from the corresponding SM
predictions are shown to be ${\cal O}(1)\%$ when the triplet and
singlet Higgs boson masses are smaller than $\simeq 1~{\rm TeV}$.
Such deviations can be measured at future electron-positron colliders.
When the masses of the MSSM-like charged Higgs boson and the MSSM-like
CP-odd Higgs boson are below $\simeq 500~{\rm GeV}$, their mass
difference is larger than the MSSM prediction by ${\cal O}(1)\%$ -
${\cal O}(10)\%$, and measurable at the LHC.  By combining these
observations, we can distinguish our model, MSSM and NMSSM.  We
emphasise that the supersymmetric grand gauge-Higgs unification model
is a good example to show capability of colliders for testing GUT
scale physics.

\begin{acknowledgments}
  The author would like to thank Shinya Kanemura,
  Hiroyuki Taniguchi and Toshifumi Yamashita for the fruitful collaboration.
\end{acknowledgments}

\end{document}